\documentclass{article}
\usepackage{datetime}
\usepackage{amsmath}
\usepackage{amssymb}
\usepackage[font=footnotesize,width=1.1\textwidth,labelfont=bf]{caption}
\usepackage[margin=1in]{geometry}

\usepackage{graphicx}
\usepackage{xcolor}

\title{Coordinated Car-following Using Distributed MPC}
\author{Di Shen, \;Qi Dai, \;Suzhou Huang\thanks{Corresponding author.}\\
Independent Researcher, \{saddieyoyo, \;daiqi5477, \;huang0suzhou\}@gmail.com}
\begin{document}
\date{}
\maketitle
 
\begin{abstract}
Within the modeling framework of Markov games, we propose a series of algorithms for coordinated car-following using distributed model predictive control (DMPC). Instead of tracking prescribed feasible trajectories, driving policies are solved directly as outcomes of the DMPC optimization given the driver's perceivable states. The coordinated solutions are derived using the best response dynamics via iterated self-play, and are facilitated by direct negotiation using inter-agent or agent-infrastructure communication. These solutions closely approximate either Nash equilibrium or centralized optimization. By re-parameterizing the action sequence in DMPC as a curve along the planning horizon, we are able to systematically reduce the original DMPC to very efficient grid searches such that the optimal solution to the original DMPC can be well executed in real-time. Within our modeling framework, it is natural to cast traffic control problems as mechanism design problems, in which all agents are endogenized on an equal footing with full incentive compatibility. We show how traffic efficiency can be dramatically improved while keeping stop-and-go phantom waves tamed at high vehicle densities. Our approach can be viewed as an alternative way to formulate coordinated adaptive cruise control (CACC) without an explicit platooning (or with all vehicles in the traffic system treated as a single extended platoon). We also address the issue of linear stability of the associated discrete-time traffic dynamics and demonstrate why it does not always tell the full story about the traffic stability.
\end{abstract}

\newpage
\section{\bf Introduction}
Designing traffic systems that possess desirable properties, such as high vehicle throughput with smooth flow, has been a lofty goal for many researchers. The task is difficult because both traffic bottlenecks and human driving behaviors, which potentially can have infinite variety, generally play intertwining roles. One common intuition is that smart traffic systems need to have some degree of coordination that goes beyond naturalistic behaviors of heterogeneous individuals. In order to push the idea of coordination to the next level, we first concentrate on the simplest case of car-following in this work, where there is no traffic bottleneck involved, but dynamic traffic jams or stop-and-go phantom waves can still form spontaneously induced purely by human driving behaviors \cite{Sugiyama, Tadaki}. In this particular setting, our task reduces to improving traffic efficiency while keeping the traffic oscillation to a minimum when vehicle density is sufficiently high.

Our approach is based on the Markov game modeling framework proposed in \cite{FrameworkPaper}. In such a framework, explicit coordination can be obtained through the concept of Nash equilibrium. Computationally, the Nash equilibrium can be well approximated using iterated best response dynamics, a specific form of self-play, which turns out to nicely coincide with the so-called distributed model predictive control (DMPC) for multi-agent systems \cite{MPCBook}. Instead of specifying the driving policy, we solve it from a pertinent MPC by maximizing the individual's objective (or utility function). In the car-following setting, only three components are included in the utility: a reward for moving forward with an ideal speed, a penalty for moving backward, and a penalty for the subjectively perceived risk of pairwise collision. We can seamlessly bridge the gap between micro-modeling (vehicle level) and macro-modeling (traffic level) by directly aggregating all the individual behaviors. No explicit platooning is necessary, in contrast with standard coordinated adaptive cruise control (CACC) approaches. The utility maximization for individual behavioral modeling also enables the central authority to search for the optimal traffic control policy at the system level offline, a game theory procedure known as mechanism design for multi-agent systems with full individual incentive compatibility.

\begin{figure}[!h]
\centering
\includegraphics[width=0.85\textwidth]{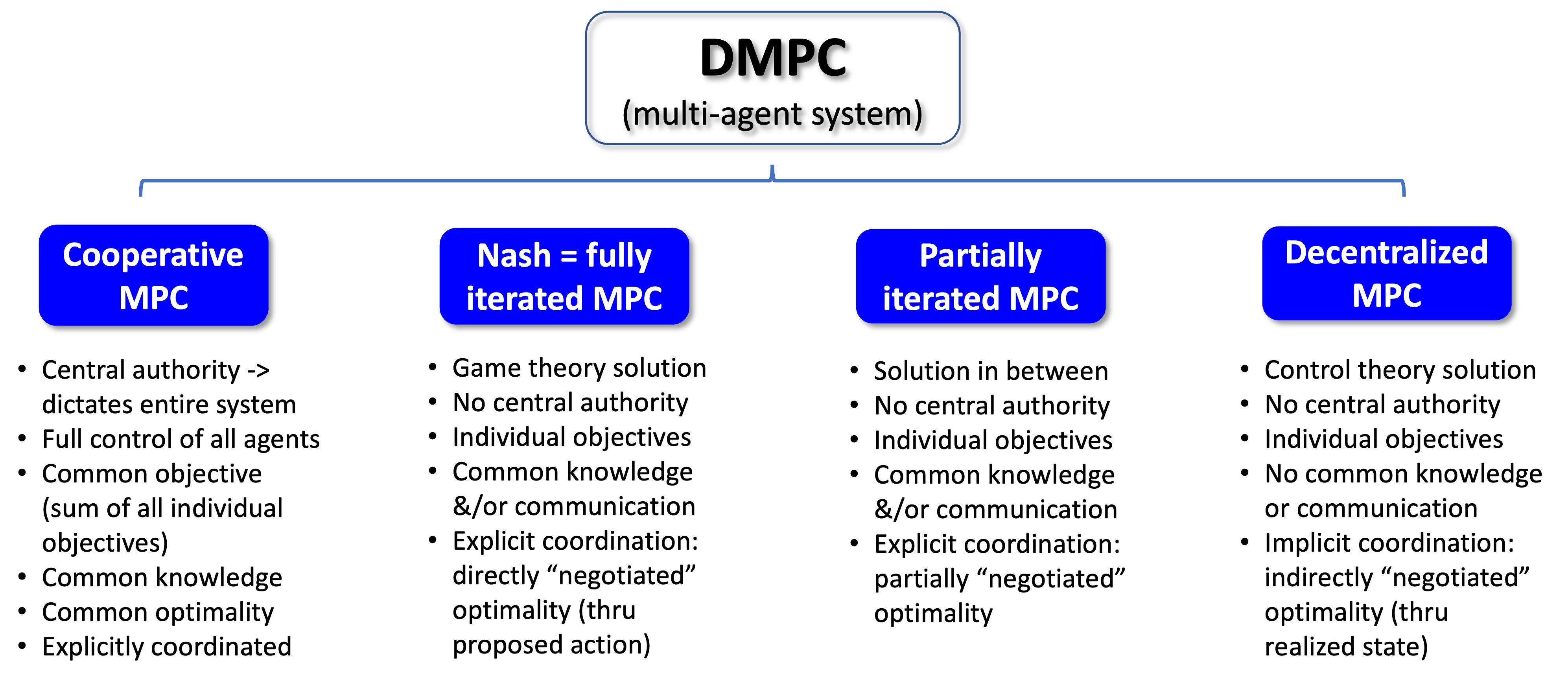}
\caption{Description of the four solution concepts under the rubric of the DMPC modeling framework. By ``No central authority'' we only mean that drivers' decision-making is not centralized. Central authority may still be needed for facilitating the coordination technically, such as providing communication channels among agents.}
\label{DMPC-Framework}
\end{figure}

For readers' convenience, we summarize all the relevant elements in formulating DMPC in Figure \ref{DMPC-Framework}.
These include whether there is a central authority, whether the objective is individualized, and whether explicit communication, negotiation, and coordination mechanism are involved. The solution concept called decentralized MPC in Figure \ref{DMPC-Framework} is closest to human driving. This in turn allows us to quantitatively estimate the model parameters using the standard statistical procedure with observed vehicle trajectory data \cite{CalibrationPaper}. The performance of the quantified model with implicit coordination can ultimately serve as the benchmark against which all the improvements brought about by explicit coordination are measured. Our final solutions are based on the partially iterated MPC, which lies conceptually between decentralized MPC (non-iterated) and Nash (fully iterated). To implement the relevant MPCs in real time, we develop several bespoke techniques so that each coordinated solutions can be derived in about 10 ms for dozens of vehicles per time step of 1/6 second on a regular laptop. Other onboard hardware and communication requirements are similar to what is required by CACC. However, computation can be performed in either a distributed manner (onboard individual vehicle) or in a centralized manner (via roadside infrastructure). If we are willing to use simulations as investigative tools, many realistic elements, such as handling arbitrary initial conditions, vehicle heterogeneity, communication latency and interruption, imperfection of perception, can be taken into account in the investigation.

Our numerical simulations show very promising results, when combined with the mechanism design of optimally tuning a single parameter in an individual's MPC objective (the ideal speed), which we assume is controllable by the traffic authority in explicitly coordinated solutions and can be derived offline. At a high vehicle density regime where stop-and-go waves form spontaneously for human driving, the coordinated solutions can improve the average speed of the fleet by as much as 100\% without suffering from oscillatory waves with large amplitude. Such improvements are possible because coordinated solutions are able to attenuate rough transients much more quickly than human driving can. In addition to improved efficiency and smoothness at the fleet level, we illustrate that better traffic safety consequently emerges among neighboring vehicles.

%
%
The remainder of the paper is organized as follows. We start with a brief review of the relevant literature and highlight the main difference from our approach. Then the modeling framework and proposed algorithms for coordinated car-following are described in the next two sections. Major results for improved traffic efficiency and smoothness are presented using simulation experiments in Section \ref{Results}. To understand the intuition behind the improvements we carry out a number of ablation studies, as well as investigate other benefits associated with reaching smoother dynamics faster, all in Section \ref{Insights}. We summarize the current work and outline some possible directions for future study in Section \ref{Summary}. We conclude with two appendices, one on the details of the utility function and its parameters and the other on the linear stability of the coordinated solutions.

\section{Related Work}

The challenge of mitigating traffic instabilities, particularly the spontaneous formation of stop-and-go waves, has become a central focus of transportation research. A growing consensus suggests that even a small penetration of automated vehicles (AVs) can act as mobile actuators to stabilize the entire traffic stream \cite{WBM18}. Foundational field experiments by Stern et al. \cite{SCM18} provided critical empirical evidence, demonstrating that the intelligent control of a single autonomous vehicle could successfully dissipate phantom traffic jams on a closed track. This finding has spurred the development of various control strategies, which can be broadly categorized into model-based, learning-based, and cooperative approaches.

Model-based strategies often leverage vehicle-to-everything (V2X) communication within specific architectures. For instance, some propose control for pairs of connected AVs (CAVs) to stabilize the human-driven vehicles between them \cite{GOM23}, while others use downstream traffic estimates for speed harmonization \cite{FKW23}. In parallel, learning-based methods, particularly Reinforcement Learning (RL), have shown considerable promise by learning effective policies directly from data. These controllers have been trained on real-world trajectory data \cite{LVN22} and have consistently outperformed other algorithms in comprehensive benchmark studies \cite{CBB22, WKP21}. However, both approaches have limitations: Model-based methods can be rigid in their communication requirements; and while RL controllers are powerful, benchmark results still show significant room for improvement \cite{CBB22}, and the resulting policies can be difficult to generalize or interpret.

Cooperative Adaptive Cruise Control (CACC) is another major research area that uses V2V communication to enable platoons of vehicles to follow each other with shorter headways, promising gains in lane capacity and stability \cite{SSL12, GAH18, SAN10, HUZ22}. The real-world effectiveness of this technology is nuanced; experimental tests show that some commercial ACC systems can be string unstable due to delays \cite{MIS14}, whereas CACC overcomes this with feedforward information \cite{MSS13}. Advanced CACC designs are now incorporating downstream traffic information to create congestion-aware algorithms \cite{KIK22}. An alternative heuristic-based strategy, Jam-Absorption Driving (JAD), involves a single vehicle proactively creating a large gap to absorb an oncoming wave \cite{NTS13, HZS16, TNE15}. A common theme in these cooperative strategies is their reliance on predefined structures, such as small, tightly-coupled platoons \cite{PLC23}, or specific, prescribed maneuvers.

The objective-driven approach in this paper, grounded in Markov games and distributed model predictive control (DMPC), offers a compelling alternative that addresses these gaps. By solving for driving policies directly from each agent's utility function, our method is more parsimonious in its parameterization and inherently more generalizable than policy-based RL. It diverges from CACC and JAD by avoiding rigid, predefined control structures. Instead, our framework treats the entire traffic stream as a single, extended platoon, where coordination is an emergent property of the DMPC optimization. This provides a natural foundation for system-level mechanism design with full incentive compatibility, offering a more flexible and scalable solution for traffic stabilization.

\section{\bf Modeling framework}
In this section we first describe the traffic setting we are dealing with in this paper. Then we formulate the modeling framework at micro-level. We conclude this section by showing how to aggregate micro-level (or vehicle-level) modeling to macro-level (or traffic-level) modeling.

\subsection{The modeling environment}
Since the stop-and-go phantom waves mostly happen within the same lane, we concentrate on environments of pure car-following without lane changes. The Tadaki setting \cite{Tadaki}, an extended version of the Sugiyama setting \cite{Sugiyama} with vehicle density dependence, is appropriate for our purposes. In this setting there are $N$ vehicles traversing on a single-lane circular road with circumference $C$ (m), yielding a vehicle density of $\rho\equiv N/C$. The vehicle ordering is chosen so that vehicle $i$ is always behind vehicle $i+1$. A periodic boundary condition is imposed, i.e. all position variables $x_{i,t}$ (and headway) should be understood as $\pmod{C}$, and the $(N+1)$-th vehicle is identified as the first vehicle.  

\subsection{Micro-level modeling}
The material in this subsection is essentially an encapsulation of what had been described in \cite{FrameworkPaper, CalibrationPaper}. However, we use new notation in order to facilitate our distributed MPC algorithms to be presented in the next section. Only the longitudinal dynamics are modeled explicitly.

There are two kinds of state variables at micro level: 1) the kinematic state that characterizes the motion of vehicle $i$: $\xi_{i,t}=\{x_{i,t},v_{i,t},a_{i,t}\}$, whose components are position, velocity and acceleration respectively; 2) agent $i$'s driving decision-making state: $s_{i,t}=\cup_{l=-\underline{m}}^{\overline{m}}\xi_{i+l,t}$\footnote{Conceptually, $\xi$'s components appearing in $s_{i,t}$ could be different from the $\xi$'s for characterizing vehicle's motion, due to drivers' estimation error. We will ignore such differences in this work for simplicity.}. Here $\underline{m}$ and $\overline{m}$ are the numbers of vehicles behind and ahead of the ego vehicle $i$, respectively, that must be attended to during decision-making. The decision variable related to the control input for driver $i$ at time $t$ is denoted by $u_{i,t}$. 

Given our current car-following setting, the kinematic state evolves according to the simple particle model: $\forall i\in\{1,2,...,N\}$,
\begin{equation}\begin{cases}
x_{i,t+1}&=x_{i,t}+v_{i,t}\Delta t+\epsilon_{i,t+1}^x 
\quad\mod{C}\\
v_{i,t+1}&=v_{i,t}+a_{i,t}\Delta t+\epsilon_{i,t+1}^v \\
a_{i,t+1}&=\gamma\, a_{i,t}+\big( u_{i,t}-\gamma\, u_{i,t-1} \big)
+\epsilon_{i,t+1}^a
\end{cases}\, ,
\label{VehicleDynamics}
\end{equation}
where $\gamma\in(0,1)$ is an AR(1) parameter for taking care of the stickiness of the vehicle dynamics (cf. \cite{CalibrationPaper}), and the noises appearing in the above state evolution are all assumed to be IID normal.

Since driving involves decision-making in a time sequence with multi-agent interactions, the most relevant modeling framework is a dynamic game. Within such a framework, one central concept is the best response defined via the following utility optimization:
\begin{equation}
\bar{u}_{i,0:H}(s_{i,t}|u_{-i,0:H})
=\displaystyle{\arg\max_{u_{i,0:H}{\in\mathcal{U}}}}
U_{i,t}^\text{eff}\big(u_{i,0:H}\big|s_{i,t},u_{-i,0:H}\big)\, .
\label{BestResponse}
\end{equation}
where $H+1$ is the planning horizon, $-i$ denotes all other agents who interact with $i$ and $\mathcal{U}=[u_\text{min},u_\text{max}]$ is the feasible action interval of the vehicle. The effective utility depends on anticipated future state sequences $\hat{s}_{i,1:H+1}$ and $\hat{s}_{-i,1:H+1}$, which in turn depend on their respective future action sequences $u_{i,0:H}$ and $u_{-i,0:H}$. In other words, the best response of agent $i$ is its optimal reaction sequence to the arbitrarily given action sequences of other agents, in addition to also being contingent on the current state $s_{i,t}$. The action sequences enter the utility function via deterministically anticipated future states $\hat{s}$ defined as:
$\forall h\in\{0,\cdots,H\}$ and current condition $\hat{s}_{i,0}=s_{i,t}$,
\begin{equation}
\begin{cases}
  \hat{x}_{i,h+1}&= \hat{x}_{i,h}+\hat{v}_{i,h}\,\Delta t \,\,\,\pmod{C}\\
  \hat{v}_{i,h+1}&=\hat{v}_{i,h}+\hat{a}_{i,h}\,\Delta t\\
  \hat{a}_{i,h+1}&=u_{i,h}
\end{cases}\quad\text{and}\quad
\begin{cases}
  \hat{x}_{-i,h+1}&= \hat{x}_{-i,h}+\hat{v}_{-i,h}\,\Delta t \,\,\,\pmod{C}\\
  \hat{v}_{-i,h+1}&=\hat{v}_{-i,h}+\hat{a}_{-i,h}\,\Delta t\\
  \hat{a}_{-i,h+1}&=u_{-i,h}
\end{cases}.
\label{AnticipationProcess}
 \end{equation}
A Nash equilibrium is achieved when all the best responses are mutually consistent with one another or when action sequences for all agents (i.e. both $i$ and $-i$) are simultaneously optimal in Eq.(\ref{BestResponse}). 

Conceptually, it is important to distinguish the two state evolutions defined in Eq.(\ref{VehicleDynamics}) and Eq.(\ref{AnticipationProcess}). The former models the mechanical realization for vehicle $i$ given driver $i$'s control inputs, whereas the latter models driver $i$'s mental anticipation of future states for all relevant agents. 

To avoid being myopic, the effective utility is typically defined by aggregating per-period utilities over the planning horizon ($H+1$).
The per-period utility is further decomposed into several components, each of which is designed to accomplish a specific purpose. Therefore, the effective utility can be written as a weighted sum
\begin{equation}
U_{i,t}^\text{eff}\big(u_{i,0:H}\big|s_{i,t},u_{-i,0:H}\big)=
\sum_{h=0}^H\sum_{k}\, w_{i,k}\,U_{i,t}^{(k)}\big(u_{i,h}
\big|\hat{s}_{i,h},u_{-i,h}\big)\Big|_{\hat{s}_{i,0}=s_{i,t}}\, ,
\label{Cumulative}
\end{equation}
where $U_{i,t}^{(k)}$ is the $k$th component of the per-period utility that depends only on the actions ($u_{i,h}$ and $u_{-i,h}$) at that period given the state of the period $\hat{s}_{i,h}$ through the anticipation in Eq.(\ref{AnticipationProcess}). 

However, there is another form of the effective utility associated with {\it adaptiveSeek}, in order to closely mimic human driving behavior \cite{CalibrationPaper}. To model the bounded rationality of human decision-making we set $u_{i,0:H}=u$ and $u_{-i,0:H}=0$, with the anticipated sequence of each per-period component transformed with some function $g_k$. Hence, this second non-standard effective utility takes the form 
\begin{equation}
U_{i,t}^\text{eff}\big(u_{i,0:H}\big|s_{i,t},u_{-i,0:H}\big)=
\sum_{k}\, w_{i,k} \,g_k\big[\displaystyle{\cup_{h=0}^H}\,U_{i,t}^{(k)}\big(u_{i,h}
\big|\hat{s}_{i,h},u_{-i,h}\big)\big] \, .
\label{g_transformed}
\end{equation}
This atypical version of the effective utility is called $g$-transformed, in contrast with the standard cumulative utility in Eq.(\ref{Cumulative}).

Lastly, we also consider centralized optimization solutions. In such cases, each agent aims to maximize the total effective utility over the entire fleet. As a consequence, the individual utility on the right-hand side of Eq.(\ref{BestResponse}) is replaced by $\sum_{i=1}^N U^\text{eff}_{i,t}$. Of course, only those individual utilities that depend on $u_{i,0:H}$ actually matter for the optimization.

The functional form of the per-period utility components $U^{(k)}_{i,t}$ and $g_k$ are given in Appendix \ref{UtilityComponents}. The associated parameters and weights are systematically calibrated via a maximum likelihood estimation of the state space model of observed vehicle trajectories from the Sugiyama experiment in \cite{CalibrationPaper}. 

\subsection{Macro-level modeling}
To bridge the gap from micro modeling to macro modeling, we merely need to stack up all kinematic states to form the macro state variable: $\bar{\boldsymbol{\xi}}_{t}=\cup_{i=1}^{N}\bar{\xi}_{i,t}\,$, once the control inputs are replaced by their corresponding equilibrium solutions (or their approximations) subject to some initial condition. It is understood that this macro state variable is from the point of view of the central traffic authority.
The traffic dynamics are then governed by the map of the following autonomous discrete-time dynamical system:
\begin{equation}
\bar{\boldsymbol{\xi}}_{t+1}=\boldsymbol{f}\big(\bar{\boldsymbol{\xi}}_{t},\,\bar{\boldsymbol{\xi}}_{t-1}\big)\, ,
\end{equation}
where the functional form of $\boldsymbol{f}$ is specified by the individual state evolution in Eq.(\ref{VehicleDynamics}). Note that this map is a second order difference equation (cf. ODE with delay from \cite{Orosz}).

\begin{figure}[!h]
\centering
\includegraphics[width=0.85\textwidth]{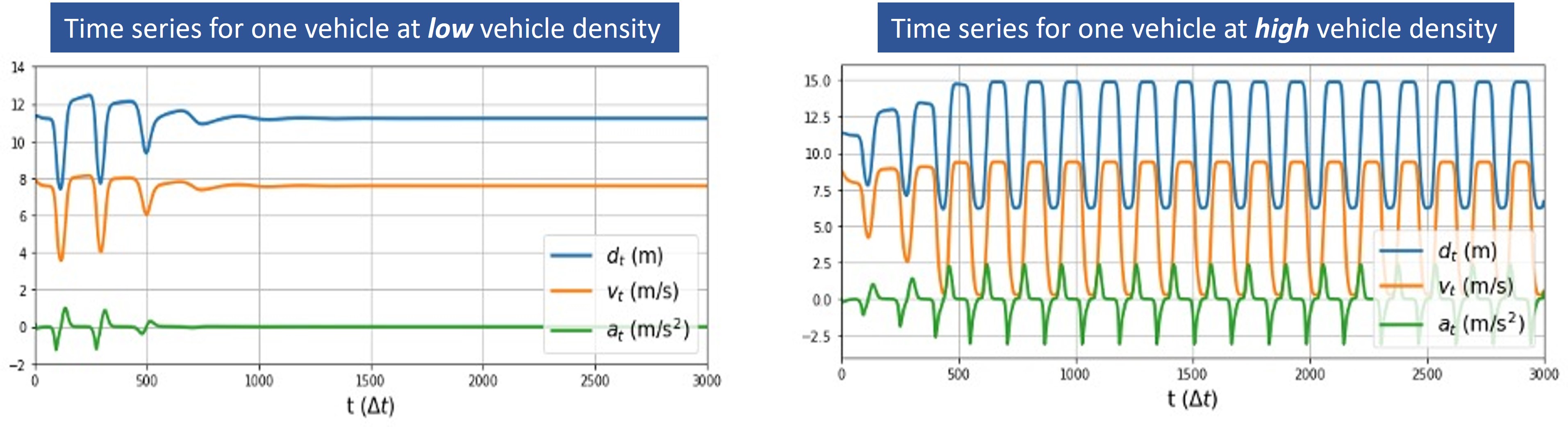}
\caption{Two possible limit solutions of the traffic dynamics when all agents are modeled as human drivers (using {\it adaptiveSeek}) with the same set of preference parameters. Left: free-flow fixed-point. Right: stop-and-go wave. The headway is defined as $d_{i,t}\equiv x_{i+1,t}-x_{i,t}\mod C$.}
\label{AsymptoticStates}
\end{figure}

Under the dynamics of {\it adaptiveSeek}, there are two limit solutions (or asymptotically stable orbits) that the traffic dynamics can converge to at a long-time limit. When the vehicle density is low, the traffic dynamics converge to a free-flow fixed point. When the vehicle density is high, the traffic dynamics converge to a stop-and-go wave.\footnote{The stop-and-go wave is a quasi-periodic orbit of the traffic dynamics, according to \cite{BifurcationPaper}} Figure \ref{AsymptoticStates} illustrates corresponding time series of the kinematic states for both solution types respectively. 
These traffic patterns closely resemble those observed in the Sugiyama \cite{Sugiyama} and Tadaki \cite{Tadaki} experiments with real human drivers.
Additional limit solutions that lie between the two extrema in Figure \ref{AsymptoticStates}, such as shallow waves, can emerge for explicitly coordinated solutions (see Appendix \ref{BifurcationAppendix} for an illustration).

To measure which asymptotic phase is prevailing, we introduce two order parameters at the traffic level. The first characterizes the traffic efficiency --- average velocity: $V\equiv\displaystyle{\text{mean}_{\substack{i,t}}\big(v_{i,t}\big)}$. The second characterizes the traffic smoothness --- oscillation amplitude: $A\equiv\text{mean}_t\big(\displaystyle{\max_i v_{i,t}-\min_i v_{i,t}}\big)$. Both time averages are done after skipping transients. For the free-flow fixed point, $A\rightarrow 0$, whereas for the stop-and-go wave, $A$ is very high. From the macro perspective, keeping both a high $V$ and a low $A$ for the traffic system is always preferred whenever possible. 

We realized in \cite{BifurcationPaper} \cite{TamingVSA} that the traffic dynamics is not only sensitive to vehicle density, but also to one of the micro-level preference parameters in the utility, the so-called ideal speed $v^*$, which is the speed that the driver would like to maintain when no one else is within the range of interaction. When the system intervention is turned on in coordinated driving settings, this special parameter effectively becomes a control input of the central traffic authority \cite{TamingVSA}. Under such situations, $v^*$ can be chosen optimally by maximizing traffic flow subject to an upper bound on oscillation amplitude (a procedure that was called computational mechanism design in \cite{TamingVSA}): 
\begin{equation}
v^*_\text{opt}(\bar{\Lambda})\equiv\arg\displaystyle{\max_{v^*}}\; V|_{A\le\bar{\Lambda}}\, .
\label{v*_opt}
\end{equation}
It is understood that the above optimization can be done offline via simulations. In an idealized situation where all agents are identical and all noises are switched off we can choose $\bar{\Lambda}=0$, which is the case we concentrate on in this paper. 

\section{Proposed algorithms}
Generally, solving for Nash equilibria in dynamic game settings is exceedingly difficult, because coupled high-dimensional optimizations are involved. Certain approximations are unavoidable to achieve the goal in real time applications. To this end, we propose a series of iterative algorithms so that the Nash equilibrium can be approximated to various degrees. They can all be viewed as some form of distributed MPC, as defined in \cite{MPCBook} and briefly summarized in the Introduction. Because multiple agents are simultaneously involved in the optimization process, all solutions concepts are coordinated, either implicitly or explicitly, whose precise meaning will be further elaborated below.

\subsection{Iterated best response dynamics}
Explicit coordination can be achieved via the iterated best response dynamics in a Markov game with repeated self-play among nearby agents (see \cite{FrameworkPaper}). It starts with the given state for ego agent $i$ and the initial MPC action sequences for all agents: $s_{i,t}=\cup_{l=-\underline{m}}^{\overline{m}}\xi_{i+l,t}$ and $\bar{u}_{i=1:N,h=0:H}^{\tau=0}=0$, where $H+1$ is the planning time horizon of the MPC (roughly a few seconds), and $\overline{m}$ and $\underline{m}$ are positive integers representing the number of vehicles ahead and behind the ego needs to attend to in decision-making. The following iterative process of the best response dynamics, dubbed $\tau$-loop, is then executed in parallel: 
$\forall i\in\{1,\cdots,N\}$,
\begin{equation}
\bar{u}^{\tau+1}_{i,0:H}(s_{i,t}|\bar{u}^\tau_{-i,0:H})
=\displaystyle{\arg\max_{u_{i,0:H}{\in\mathcal{U}}}}
U_{i,t}^\text{eff}\big(u_{i,0:H}\big|s_{i,t},\bar{u}^\tau_{-i,0:H}\big)\, .
\label{DMPC}
\end{equation}
Note that when solving the action sequence for agent $i$ at step $\tau+1$ in the above equation, the action sequences for all other agents at step $\tau$ are treated as given and fixed. More explicitly, the anticipated future states in Eq.(\ref{DMPC}) for agent $i$ and agent $-i$ evolve respectively according to, $\forall h\in\{0,\cdots,H\}$,
\begin{equation}
\begin{cases}
  \hat{x}_{i,h+1}= \hat{x}_{i,h}+\hat{v}_{i,h}\,\Delta t\,\,\, \pmod{C}\\
  \hat{v}_{i,h+1}=\hat{v}_{i,h}+\hat{a}_{i,h}\,\Delta t \\
  \hat{a}_{i,h+1}=u_{i,t}
 \end{cases}\text{and}\quad
\begin{cases}
  \hat{x}_{-i,h+1}= \hat{x}_{-i,h}+\hat{v}_{-i,h}\,\Delta t\,\,\, \pmod{C}\\
  \hat{v}_{-i,h+1}=\hat{v}_{-i,h}+\hat{a}_{-i,h}\,\Delta t\\
  \hat{a}_{-i,h+1}=\bar{u}^\tau_{-i,t}
 \end{cases} ,
 \label{Anticipation_DMPC}
 \end{equation}
which implies that the anticipation process is such that the ego agent $i$ knows its own control input in anticipating its own future state, while using other agents' action sequences from previous $\tau$-step to anticipate their future states. Note further that there are neither noises nor auto-regressive elements in the MPC's anticipation process, in contrast with the state evolution of the vehicle in Eq.(\ref{VehicleDynamics}). Because of the simultaneous execution for all agents, Eq.(\ref{DMPC}) is called a distributed MPC. 

After a pre-determined number of iterations, $T$, the optimal control input for agent $i$ at time $t$ is given by the first action in the optimization sequence:
\begin{equation}
\bar{u}_{i,t}(s_{i,t})=
\bar{u}_{i,0}^{T+1}(s_{i,t}|\bar{u}_{-i,0:H}^{T}),\quad\forall i\in\{1,\cdots,N\} \, .
\end{equation}
When $T=0$ the above algorithm is identical to {\it adaptiveSeek}  proposed in \cite{CalibrationPaper}, provided that the $g$-transformed utility is used and $u_{i,h}$ is a constant in $h$ and $u_{-i,h}=0$. When $T>0$, a communication mechanism is needed to share the previous action sequences from other interacting agents to agent $i$, for anticipation in Eq.(\ref{Anticipation_DMPC}).

\subsection{Simplifying the optimization}
The optimization problem in Eq.(\ref{DMPC}) is still too hard to solve for real time applications with a very large number of agents. To speed up the calculation we deploy the following two simplifications.
\subsubsection{Re-parameterization of control input}
The first simplification is motivated by regarding the action sequence $u_{i,0:H}$ as a curve in $\forall h\in\{0,1,\cdots,H\}$. Then we can express this curve in terms of a polynomial basis, such as
\begin{equation}
u_{i,h}(u_i^{(0:H)})=u_i^{(0)}+u_i^{(1)}\,(h\Delta t)
+\cdots+u_i^{(H)}\,(h\Delta t)^H,\quad\quad\forall h\in\{0,\cdots,H\}\, .
\end{equation}
Note that the degree of freedom is the same in either set of the optimization variables: $H+1$. The advantage of the above re-parameterization is that $u_i^{(0:H)}$ can better facilitate truncating the optimization space. Furthermore, the feasibility constraints imposed on the original MPC variables in Eq.(\ref{DMPC}) will continue to be linear in terms of the re-parameterized optimization variables. For example, the 0th order polynomial (or 1D optimization) is:
\begin{equation}
u_{i,h}(u_i^{(0)})=u_i^{(0)},\;\;\;\;\;\;\forall h\in\{0,\cdots,H\}\, ,
\end{equation}
which corresponds to the treatment of $u_{i,0:H}$ as a constant in the MPC.
The 1st order polynomial (or 2D optimization) is:
\begin{equation}
u_{i,h}(u_i^{(0:1)})=u_i^{(0)}+u_i^{(1)}\,(h\Delta t),\;\;\;\;\;\;\forall h\in\{0,\cdots,H\}\, ,
\end{equation}
which corresponds to the treatment of $u_{i,0:H}$ as a linear function of $h$ in the MPC. 

\subsubsection{Optimization: softmax via grid search}
When the dimension of the optimization space is sufficiently reduced (from $(H+1)$-dimensional $u_{i,0:H}$ to $p$-dimensional $u_i^{(0:p)}$ with $p\ll H$), the optimization can be done using computationally efficient grid searches. In so doing, we can also avoid using any fancy online non-convex solver, which could potentially get stuck at local optima. To restore the continuous nature of the action, we use the following Boltzmann average (or softmax):
\begin{equation}
\bar{u}_{i,0:H}(s_{i,t}|u_{-i,0:H})
\longrightarrow
\displaystyle{\sum_{u_i^{(0:p)}\in\mathcal{G}}}u_{i,0:H}(u_i^{(0:p)})
P_{i,t}\Big(U_{i,t}^\text{eff}\big(u_{i,0:H}(u_i^{(0:p)})\big|s_{i,t},u_{-i,0:H}\big)\Big)\, ,
\label{softmax}
\end{equation}
with $P_{i,t}\propto\exp[\lambda\, U^\text{eff}_{i,t}]$, and $\mathcal{G}$ being a reasonably dense grid on $\mathcal{U}$. When $\lambda$ is positive and large, the softmax function becomes equivalent to grid search on $\mathcal{G}$.  Thus, we can regard the optimal control input in Eq.(\ref{softmax}) as a regularized version of Eq.(\ref{BestResponse}),  such that $\bar{u}$ has a continuous support and is differentiable with respect to model parameters. Furthermore, the feasibility constraints on $u_i^{(0:p)}$ in Eq.(\ref{softmax}) can be efficiently implemented using a pre-determined mask on $\mathcal{G}$.

\subsection{Comments on the proposed algorithms}
\begin{figure}[!h]
\centering
\includegraphics[width=0.80\textwidth]{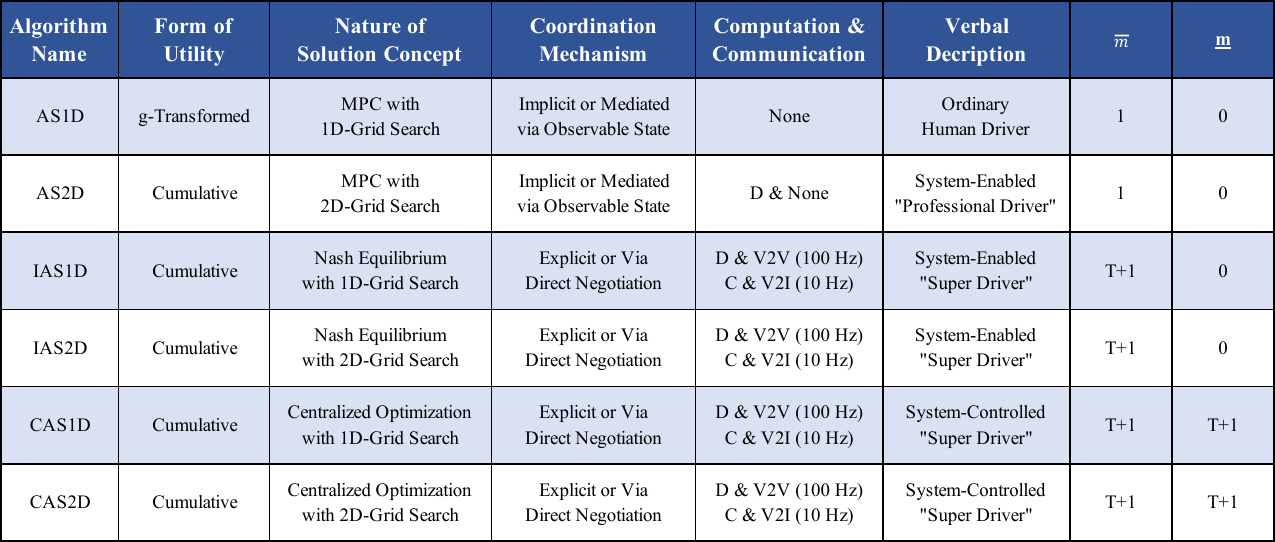}
\caption{Brief description of the proposed algorithms. In the ``Computation \& Communication'' column,``D'' stands for decentralized, i.e. computation done by the onboard computer of each vehicle, and ``C'' stands for centralized, i.e. computation done by the centralized server.}
\label{DescriptionTable}
\end{figure}

Table \ref{DescriptionTable} summarizes the six algorithms we wish to investigate, along with some of the detailed requirements and relevant verbiages. We do this by column in turn. 

{\it Algorithm Name}: The following conventions are adopted: ``AS'' for {\it adaptiveSeek}, ``1D'' for 1D-grid search, ``2D'' for 2D-grid search, ``I'' for iterated, and ``C'' for centralized. Using a subscript we also indicate which type of effective utility function is used.

{\it Form of Utility}: The $g$-transformed is only for AS1D, as it is intended for mimicking human driving in {\it adaptiveSeek} \cite{CalibrationPaper}. All others deploy the standard cumulative utility, as these are related to driving decisions made by automated vehicle or by the central control system, not a human driver. 

{\it Nature of Solution Concept\footnote{Strictly speaking, Nash equilibrium and centralized optimization solutions require $T\rightarrow\infty$. Our interest here is for finite $T$. We will later show numerically that the convergence to their limits is very quick, even for $T=2$. Therefore, to avoid any additional notational complication, we will refer the finitely iterated solutions as their corresponding limits of Nash equilibrium and centralized optimization solutions.}}: There are three solution concepts involved: distributed MPC without the $\tau$-loop iteration (AS1D and AS2D), Nash equilibrium-based (IAS1D and IAS2D), and centralized optimization based (CAS1D and CAS2D). The $\tau$-loop iterations only apply to the last two concepts.

{\it Coordination Mechanism}: When we call a coordinated solution implicit ($T=0$), it is meant that the coordination is mediated by perceivable state through the driver's own perception without explicit inter-agent communication. In such cases the decision-makers can only react to what had happened and reflected in their perception in the next time period. For explicit coordination ($T>0$) driving decisions are made via direct negotiation, in the sense that agents share among themselves what they intended to do within a single time period. 

{\it Computation \& Communication}: A detailed communication mechanism depends on where the computation is done. In the decentralized case, computations are done in each individual vehicle's onboard computer. Because direct negotiation requires sharing the intended action with their neighbors multiple times within a single time period, the inter-agent communication (or V2V) has to be done at a high frequency, say at the level of 100 Hz. In the centralized case, computation is done in a central server, and only the initial conditions and final solutions of the period need to be communicated from each vehicle to the central server (or V2I), say at the level of 10 Hz. Because the direct negotiation can happen inside the central server, high frequency inter-agent communication is avoided in the centralized computation.

{\it Verbal Description}: This is meant to provide a simple analogy relative to the baseline case of human driving (AS1D). We call AS2D ``Professional Driver'' because it requires fancier computations than a human driver can ordinarily achieve. We use ``Super Driver'' to mean that solutions with direct negotiation enabled by the central system are beyond ``Professional Driver''.

{\it $\overline{m}$ and $\underline{m}$}: These columnns indicate the range of interaction of the ego vehicle $i$ with its neighboring vehicles. $\overline{m}$ (or $\underline{m}$) is the number of vehicles ahead (or behind) that the ego needs to attend to. From the optimization perspective of the ego, the reward for moving forward or penalty for moving backward for other vehicles is not relevant. Therefore, only the pairwise collision terms enter the ego's MPC in Eq.(\ref{DMPC}). Note that, since every $\tau$-step iteration brings one extra vehicle (ahead or beyond) into the ego's attention set,  $\overline{m}$ and $\underline{m}$ depend on the depth of the $\tau$-loop. We assume that the effective utility for the central system is the sum of all individual utilities. This implies that the ego also needs to pay attention to collision risks of nearby vehicles in the rear (i.e. $\underline{m}>0$). Even though the last two algorithms, CAS1D and CAS2D, are conceptually based on centralized optimization, we continue to use the iterative method to obtain the optimality by only modifying the objective function from IAS1D and IAS2D. Otherwise, optimizing all the decision variables simultaneously would be too difficult to achieve, in either computational speed or sub-optimality.

\section{Results}\label{Results}
\begin{figure}[!h]
\centering
\includegraphics[width=0.85\textwidth]{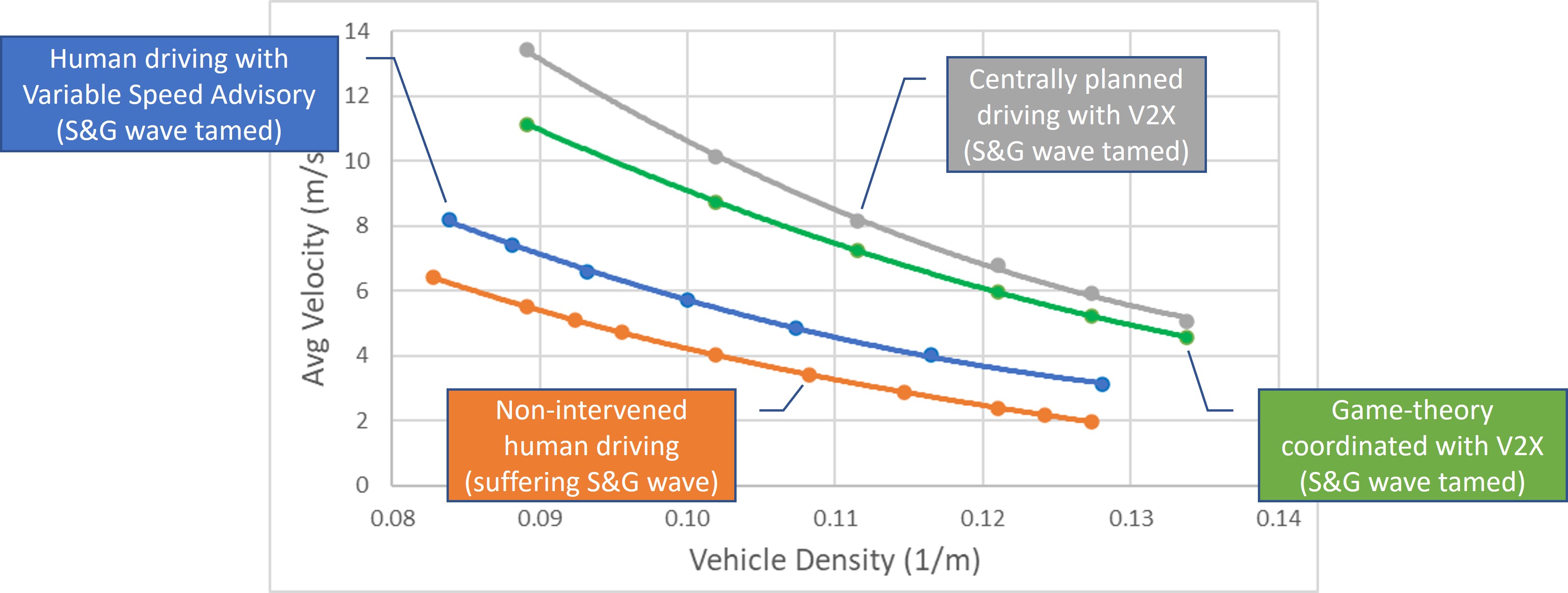}
\caption{Traffic efficiency improvement: average fleet velocity vs. vehicle density. Orange: non-intervened human driving which suffers from stop-and-go waves. Blue: human driving with a density-dependent variable speed advisory. Green: game theory coordinated solution with V2X communication. Gray: centralized optimization solution. Stop-and-go waves are suppressed in the last three cases.}
\label{BenefitPlot}
\end{figure}

We now present our main findings summarized in Figure \ref{BenefitPlot}: long-term average velocity of the entire fleet vs. vehicle density. We deliberately choose the range of vehicle density sufficiently high so that stop-and-go phantom waves are the dominant asymptotic solution if all vehicles are driven by human drivers. The baseline (orange) is the result for non-intervened human driving (AS1D$\_$g). The second curve (blue) is when the central system imposes a density dependent speed limit to all human drivers, also known as variable speed advisory (VSA) (see \cite{TamingVSA} for details). Because the speed limit is set according to Eq.(\ref{v*_opt}) with $\bar{\Lambda}=0$, there are no stop-and-go waves present in the traffic. Even with such a simple intervention mechanism the improvement of the traffic flux is non-trivial. 

On the other hand, explicitly coordinated solutions can do much better. The third case (green) is the result for the Nash equilibrium based solution IAS2D$\_$c. The fourth case (gray) is the result for the centralized optimization solution CAS2D$\_$c. Under explicit coordination, the traffic flow dramatically improves (roughly doubling that with non-intervened human driving) and stop-and-go waves are tamed at the same time. Again, the optimal ideal speed, $v^*_\text{opt}$, is set according to Eq.(\ref{v*_opt}) in these last three cases. It is worth noting that, although this is the best result in terms of traffic efficiency point of view, the incremental improvement of the centralized optimization solution relative to the Nash equilibrium based solution is not very large. Furthermore, we find the former is less robust and harder to compute than the latter, likely reflecting the fact that more agents (twice as many) are involved in the coordination process for the former than for the latter.

\section{Insights}\label{Insights}
To understand why it is possible to achieve such dramatic improvements in Figure \ref{BenefitPlot}, we first perform ablation studies to highlight the three major upgrades from the implicitly coordinated {\it adaptiveSeek} solution to fully explicit coordinated solutions: from 1D-grid search to 2D-grid search, from the $g$-transformed utility to cumulative utility, from indirect negotiation to direct negotiation. These studies can be roughly summarized in one sentence: the resulting traffic system can reach the free-flow fixed point much more quickly. We then explore 1) how smoother traffic dynamics at the micro level manifests in traffic conditions at the macro level, and 2) implications to central system control.

\subsection{Ablation studies}
In this subsection we carry out a number of paired contrasting investigations. For each pair we vary one element of modeling at a time, while keeping everything else equal. We then choose the best signal to illustrate the difference. In this way we can observe how the performance of the implicit coordinated solution with human driving (i.e. {\it adaptiveSeek}) improves progressively towards explicit coordinated solutions. From Table \ref{DescriptionTable} we can see there are three major elements involved, which we examine in turn. For a certain parameter regime, it is known that both types of asymptotically stable solutions, stop-and-go waves and free-flow fixed points, can co-exist \cite{Orosz, TamingVSA}, depending on the initial traffic condition. So, we define two qualitatively different initial traffic conditions. One is closer to the free-flow fixed point: $\{d_{i,0}=C/N, v_{i,0}=v^*-1, a_{i,0}=0\}\;\;\forall i\in\{1,\cdots,N\}$. The other is closer to stop-and-go wave, obtained from the first one by applying a kick (constant braking of $-1$ (m/s$^2$) to vehicle $N$ for 6 seconds consecutively, as long as its velocity is positive).

\subsubsection{2D-grid search $\succ$ 1D-grid search}
\begin{figure}[!h]
\centering
\includegraphics[width=0.85\textwidth]{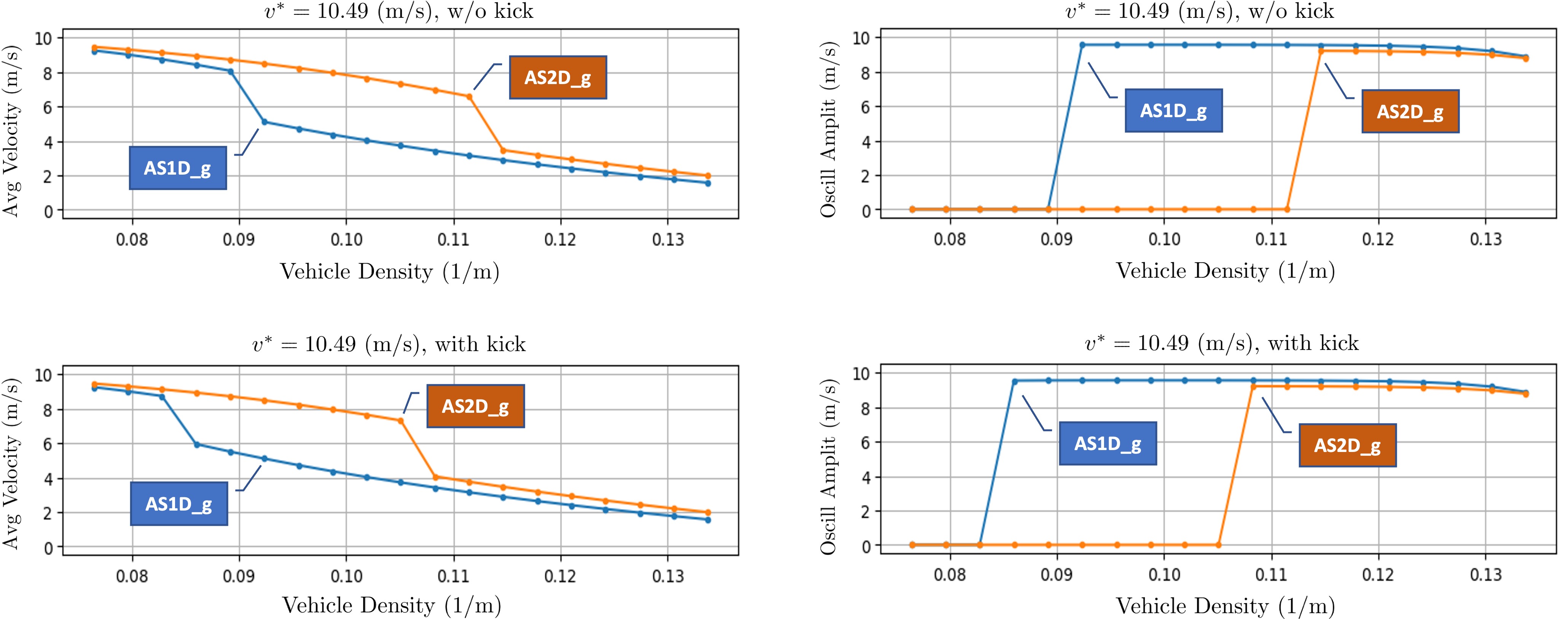}
\caption{The onset of stop-and-go waves is substantially delayed till much higher vehicle density if the grid search is upgraded from 1D to 2D in the MPC. Top row: Average velocities when the initial traffic condition is closer to the free-flow fixed point. Bottom row: Average Velocities when the initial traffic condition is closer to the stop-and-go wave.}
\label{2D_1D}
\end{figure}

In {\it adaptiveSeek}, due to humans' limited computational power, 1D-grid search is used for MPC optimization ($\text{AS1D}\_$g). When going beyond human driving, we can upgrade the MPC optimization by using a 2D-grid search ($\text{AS2D}\_$g). Figure \ref{2D_1D} indicates that this upgrade alone can substantially delay the formation of stop-and-go waves till much higher vehicle density. In the figure, we show the average velocity and oscillation amplitude for two initial conditions, one closer to the free-flow fixed point (upper row: w/o kick) and the other closer to the stop-and-go wave (lower row: with kick). While the onset of the stop-and-go wave is different, the amount of improvement is similar in both cases.

\subsubsection{Cumulative utility $\succ$ $g$-transformed utility}
\begin{figure}[!h]
\centering
\includegraphics[width=0.85\textwidth]{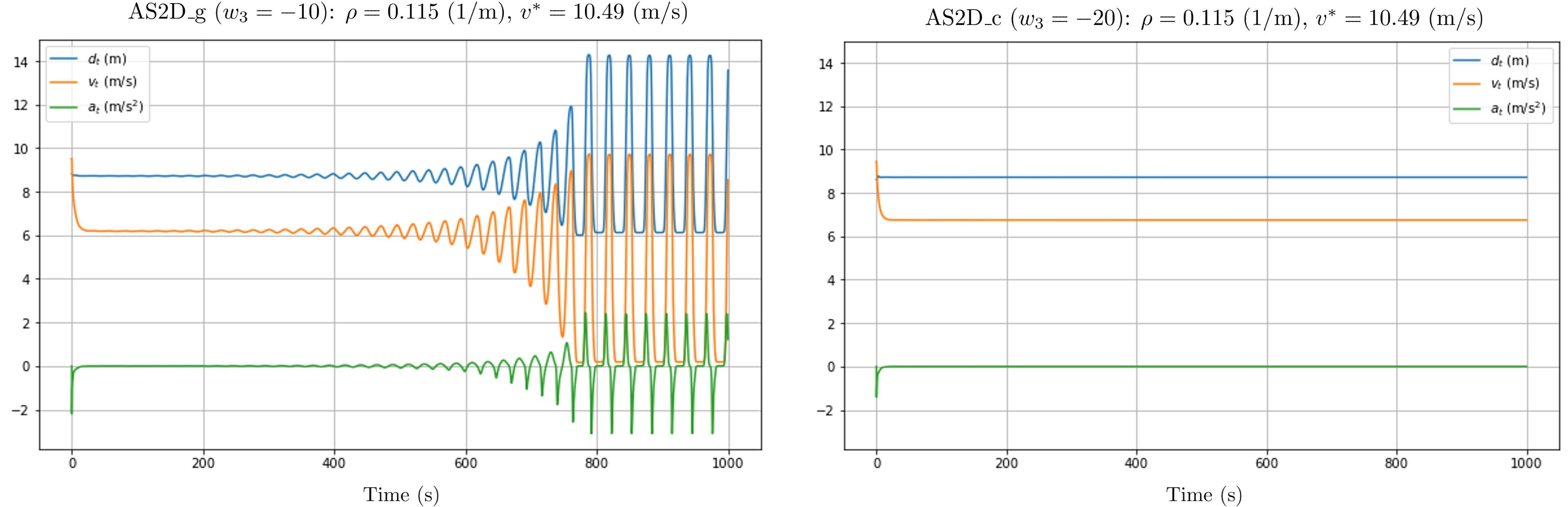}
\caption{Cumulative utility can facilitate better coordination temporally within the MPC horizon. Left: Kinematic states when the $g$-transformed utility is used. Right: Kinematic states when the cumulative utility is used.}
\label{CU_gTU}
\end{figure}

Once the vehicles are under system control, we can also replace the $g$-transformed utility in {\it adaptiveSeek} by the standard cumulative utility. In Figure \ref{CU_gTU} we show the trajectories for three state variables of one vehicle: headway, velocity, acceleration. The traffic initial conditions are the same and closer to the free-flow fixed point. On the left, the $g$-transformed utility is used ($\text{AS2D}\_$g), and on the right the cumulative utility is used ($\text{AS2D}\_$c), though both are done with 2D-grid search. Note that not only does $\text{AS2D}\_$g suffer from stop-and-go waves asymptotically, but also its long term average velocity is lower than that of $\text{AS2D}\_$c at the same vehicle density of $\rho=0.115$ (1/m).

\subsubsection{Direct negotiation $\succ$ indirect negotiation}
\begin{figure}[!h]
\centering
\includegraphics[width=0.85\textwidth]{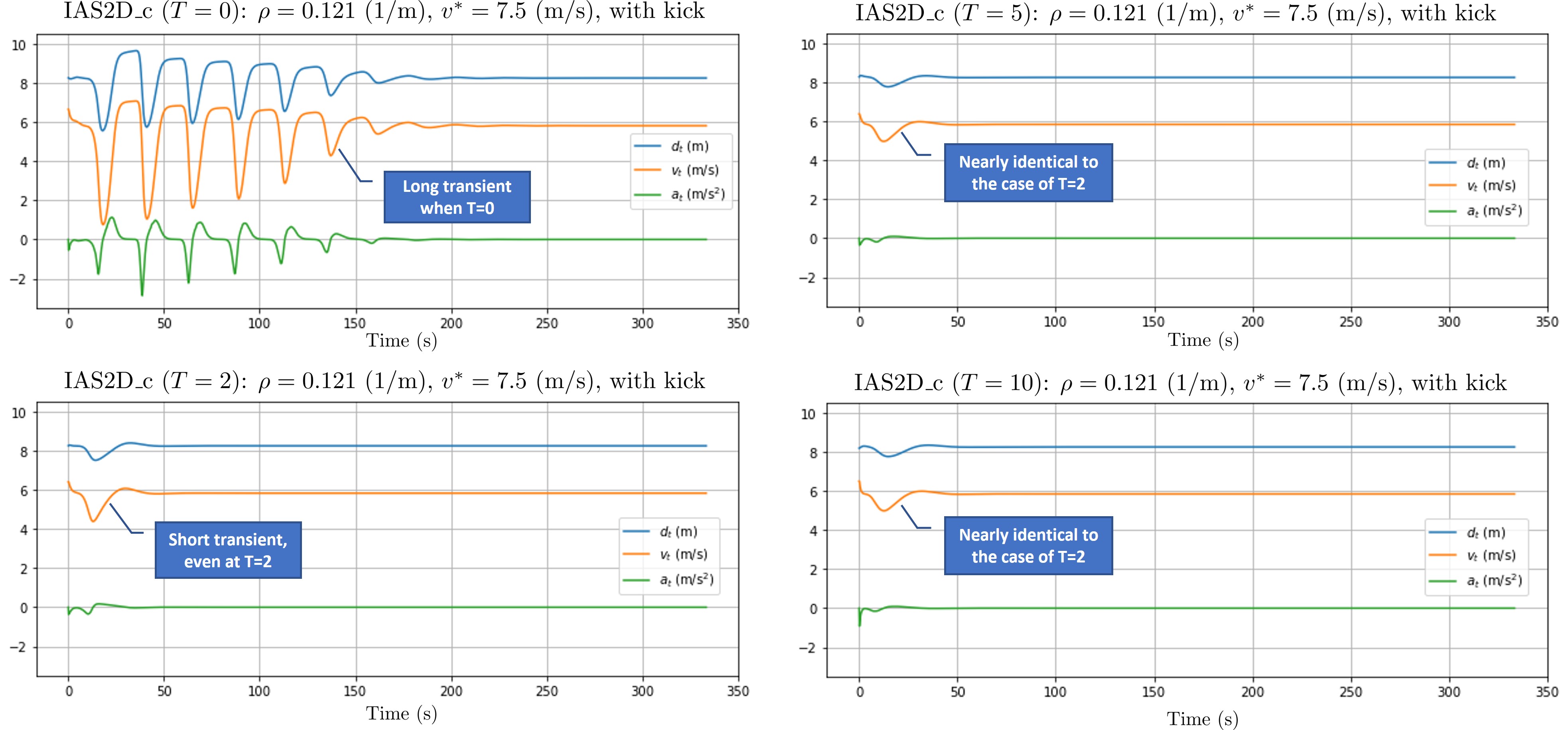}
\caption{Time series of the kinematic state for one vehicle: Faster attenuation of transients of the explicitly coordinated solutions ($T>0$) than the implicitly coordinated solution ($T=0$), and fast convergence of the $\tau$-loop.}
\label{SmoothDynamics}
\end{figure}

Explicit coordination is achievable only with direct negotiation among agents at a high frequency, say 100 Hz. This is because it is necessary for all agents to share their intended actions multiple times within a single time period at a sub-second level. The information sharing mechanism can be flexible, either directly (V2V) or via the infrastructure (V2I).
Figure \ref{SmoothDynamics} shows that the explicitly coordinated DMPC solution ($T>0$) can attenuate transients much faster than the implicitly coordinated DMPC solution ($T=0$). Furthermore, the same figure also demonstrates that the convergence of the iterated best response dynamics is very fast. Typically, it is sufficient even for $T=2$. 

\begin{figure}[!h]
\centering
\includegraphics[width=0.85\textwidth]{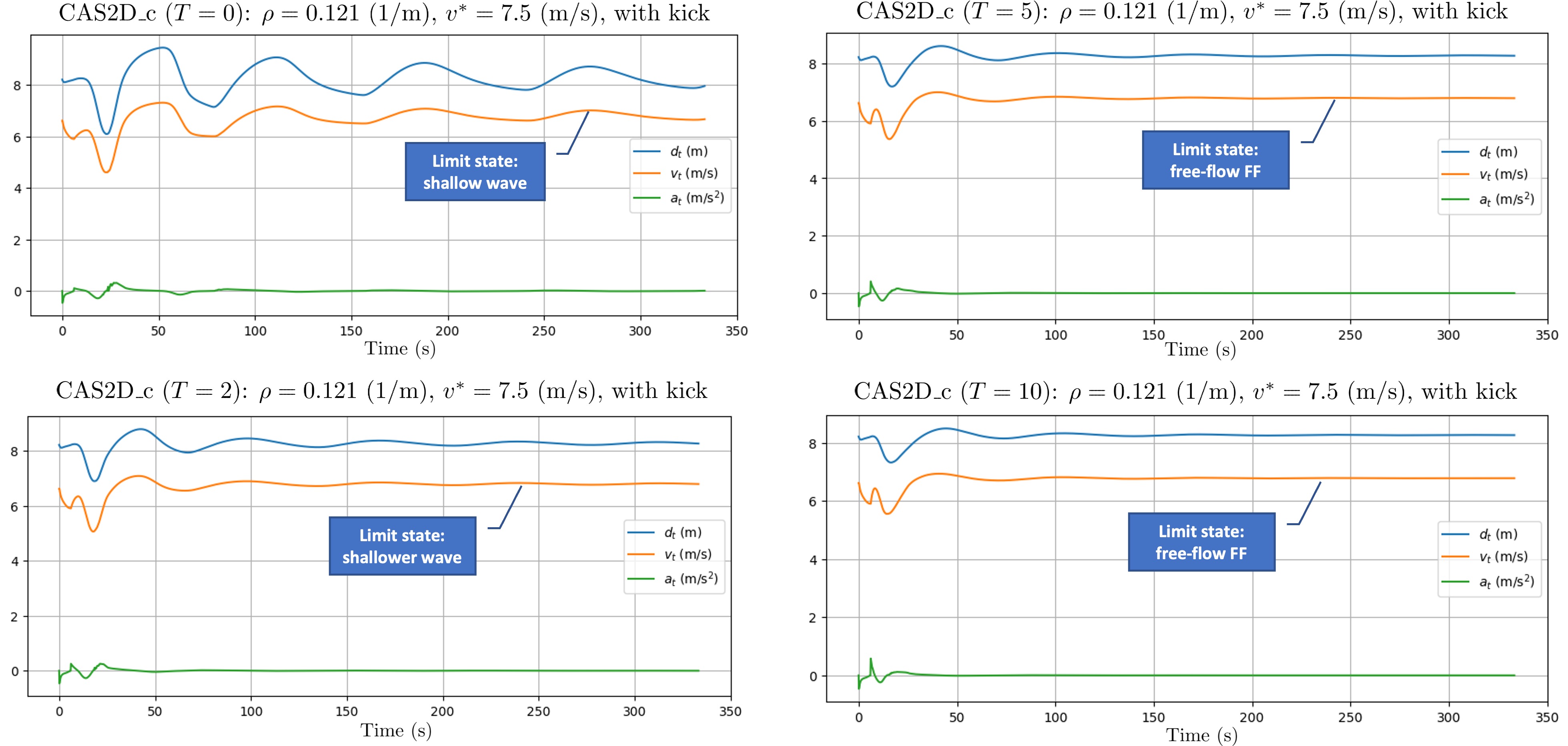}
\caption{Time series of the kinematic state for one vehicle: The $\tau$-loop for the centralized optimization solution converges also quite fast.}
\label{TauLoop2C}
\end{figure}

The fast convergence of the $\tau$-loop for iterated best response dynamics carries over to the $\tau$-loop of centralized optimization solution, as illustrated in Figure \ref{TauLoop2C}. On the other hand, subtle differences do exist between CAS2D and IAS2D. For example, the ability to attenuate transients for the centralized optimization solution is not as good as that of the iterated best response dynamics. However, the asymptotic velocity is higher in CAS2D than that in IAS2D. The general conclusion seems to be that CAS2D (interacting with vehicles both ahead and behind) can coordinate better but takes longer, whereas IAS2D (interacting with vehicles only ahead) can reach its asymptotic state quicker but with the expense of slightly lower asymptotic velocity.

Another interesting phenomenon, perhaps less appreciated phenomenon in the prior literature, is that the asymptotic states for CAS2D$\_$c with $T=0$ and $T=2$ are neither a free-flow fixed point nor a stop-and-go wave. Instead, they are very shallow waves that are nearly indistinguishable in all practical purposes from free-flow fixed point solutions, especially when noise for the state evolution is added. We will elaborate this point more in Appendix \ref{BifurcationAppendix} when we quantitatively address the linear stability issue at the traffic level.

\subsubsection{Various coordination processes}
So far we have been mostly presenting results focusing on each individual agent. However, coordination, implicit or explicit, requires participation of all agents, not just those nearby. Therefore, to truly appreciate how various coordinated solutions manifest themselves we need to examine the action sequences for all agents involved simultaneously. The same traffic initial condition as in Figure \ref{SmoothDynamics} and Figure \ref{TauLoop2C}: $\rho=0.121$ (1/m) and $v^*=7.5$ (m/s) with a kick can serve this purpose well. The kick provides a big disturbance for the entire system to respond to. Although all the asymptotic states are ultimately free-flow fixed points or very shallow waves, we are interested in how the response processes differ among the four types of qualitatively distinct coordinated algorithms.
\begin{figure}[!h]
\centering
\includegraphics[width=0.85\textwidth]{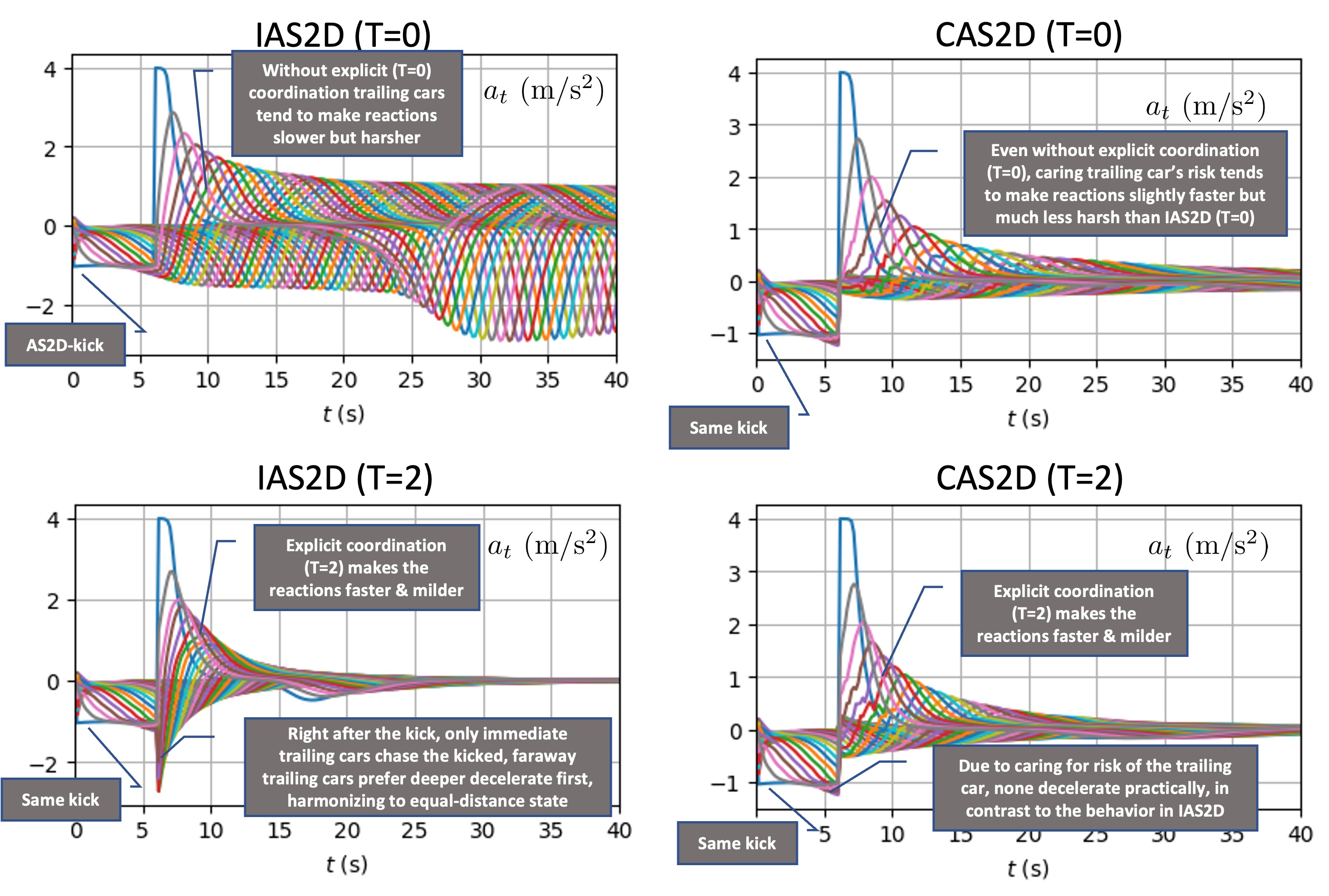}
\caption{Action sequences for all agents (blue is the kicked vehicle): Top Left: implicitly coordinated IAS2D$\_$c with $T=0$; Bottom Left: explicitly coordinated IAS2D$\_$c with $T=2$; Top Right: centralized optimization CAS2D$\_$c with $T=0$; Bottom Right:  centralized optimization CAS2D$\_$c with $T=2$.}
\label{Totality}
\end{figure}

The top-Left panel of Figure \ref{Totality} shows the action sequence for IAS2D$\_$c with $T=0$. Because there is no communication in this case, no way for agents to quickly perceive each other's intentions. Therefore, they cannot respond to other agents' actions until the realized actions are reflected in the perceivable state in the next period. This is why the implicitly coordinated solution takes a much longer time to reach its asymptotic state. On the other hand, explicitly coordinated solutions can do much better, because the driving intentions are shared via the communication mechanism, as shown by the action sequences of IAS2D$\_$c with $T=2$ in the bottom-Left panel of Figure \ref{Totality}. 

The top-Right panel of Figure \ref{Totality} shows the action sequence for CAS2D$\_$c with $T=0$, which differs from IAS2D$\_$c with $T=0$ in that each agent will try to avoid collision with not only the vehicle immediately ahead, but also the vehicle immediately behind. Attending to the rear collision risk makes the implicit coordination better, responding slightly faster and much less harshly than under IAS2D$\_$c with $T=0$. Adding communication in CAS2D$\_$c with $T=2$ makes the coordination explicit and better still, as demonstrated in the bottom-right panel of Figure \ref{Totality}.

A comparison of the bottom two panels in Figure \ref{Totality} reveals the time/nature of the difference between the explicitly coordinated IAS2D$\_$c and CAS2D$\_$c. It mostly happens right after the end of the kick process (at the 6th second), at which time the kicked vehicle (blue) accelerates maximally in order to narrow the gap with the leading vehicle, which has become big after six seconds of braking. Only the immediately trailing vehicles also follow the kicked vehicle in IAS2D$\_$c ($T=2$). Trailing vehicles further behind the kicked vehicle prefer to decelerate first (see the pronounced dip at the 6th second mark in the bottom-Left panel), so as to harmonize to the equal-distanced steady state more quickly. There is no corresponding deceleration behavior in CAS2D$\_$c, because the deceleration increases rear collision risk.

\subsection{Benefits of reaching smoother dynamics faster}
The ablation studies indicate that solutions with iterated best response dynamics (IAS2D and CAS2D) can generally reach smoother traffic dynamics faster at macro level. This in turn has a number of important practical implications.

\subsubsection{Higher $v^*_\text{opt}$}\label{higher_v*}
\begin{figure}[!h]
\centering
\includegraphics[width=0.85\textwidth]{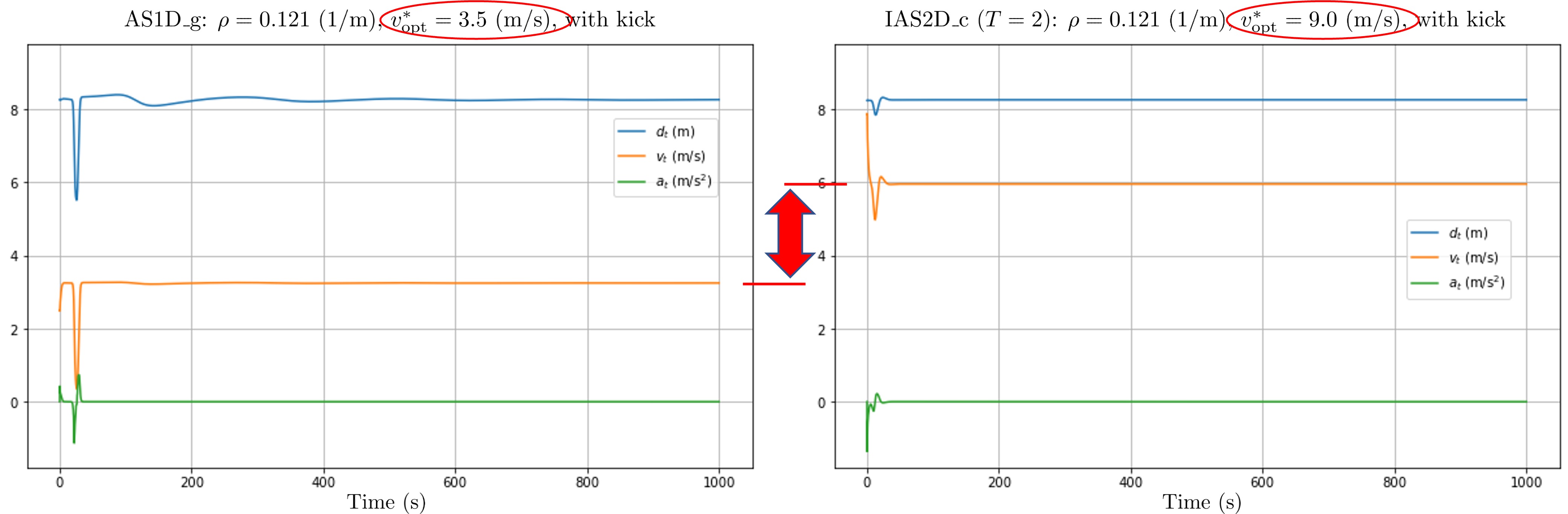}
\caption{Reaching smoother dynamics faster enables the dynamics to sustain higher $v^*_\text{opt}$ without suffering from stop-and-go waves, and ultimately higher traffic flow.}
\label{Higher_v*opt}
\end{figure}

The first obvious benefit of smoother traffic dynamics achieved via explicit coordination, as already alluded to in Section \ref{Results}, is that the traffic dynamics can sustain higher optimal ideal speed at the same vehicle density. Figure \ref{Higher_v*opt} illustrates this well. On the left, we have $v^*_\text{opt}=3.5$ (m/s) for AS1D$\_$g (a point on the blue curve in Figure \ref{BenefitPlot}), and on the right, we have $v^*_\text{opt}=9.0$ (m/s) for IAS2D$\_$c (a point on the green curve in Figure \ref{BenefitPlot}), with both at fairly high vehicle density at $\rho=0.121$ (1/m) and the same kicked initial traffic condition. While neither have stop-and-go waves, the asymptotic average speeds can differ by nearly $3$ (m/s), implying dramatically improved traffic flow.

Another interesting aspect is worth mentioning. One might naively expect that traffic dynamics are rougher under AS1D$\_$g than that under AS2D$\_$c. However, Figure \ref{Higher_v*opt} (Left: AS1D$\_$g) seems to indicate the contrary when comparing to Figure \ref{SmoothDynamics} (Top Left: IAS2D$\_$c with $T=0$, which is equivalent to AS2D$\_$c). The apparent paradox is resolved if we notice the big difference between the two $v^*$s involved, with the former at 3.5 (m/s) and the latter at 7.5 (m/s). It is intuitive to imagine that transients are harder to propagate if $v^*$ is lower, which implies that there is less tendency to chase the vehicle ahead. Even though lower $v^*$ can tame stop-and-go waves, explicit coordination can do the same much more efficiently in terms of improving traffic flow.

\subsubsection{Safer traffic}
A somewhat less obvious but otherwise equally important consequence is that smoother traffic dynamics also implies improved safety. While it is beyond the scope of this work to systematically quantify the safety improvement,\footnote{The utility component of the perceived subjective risk of pairwise collision (see Appendix \ref{UtilityComponents}) can naturally serve as a quantitative risk measure in our framework, as it is the risk penalty that underlies drivers' decision-making.} we can at least qualitatively illustrate those aspects of traffic dynamics that are intuitively important to safety improvement.
\begin{figure}[!h]
\centering
\includegraphics[width=0.88\textwidth]{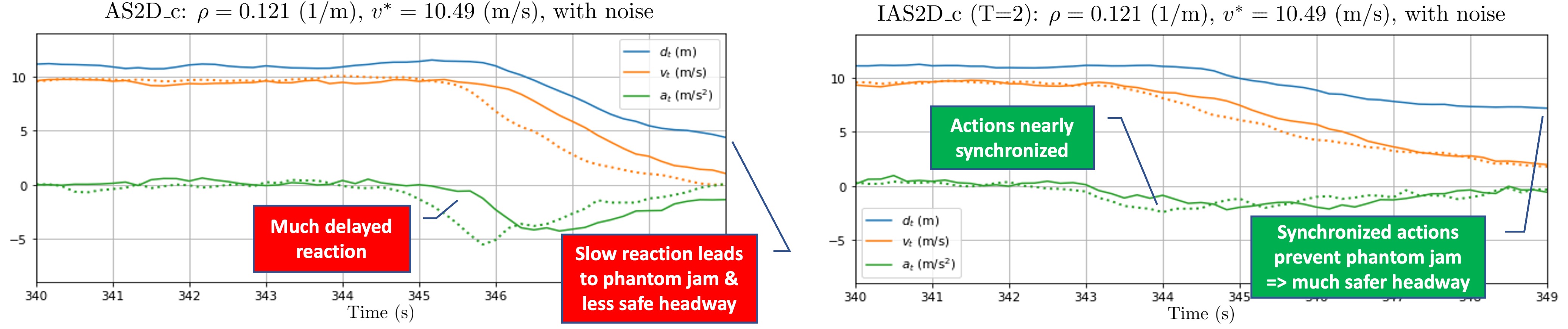}
\caption{Reaching smoother dynamics faster underlies safer traffic. Kinematic states of the ego (full line) and its leading vehicle (dotted line). Left: AS2D$\_$c. Right: IAS2D$\_$c.}
\label{SaferTraffic}
\end{figure}
We choose to contrast the traffic dynamics of AS2D$\_$c (or equivalently IAS2D$\_$c with $T=0$) and IAS2D$\_$c with $T=2$ at $\rho=0.121$ (1/M) and $v^*=10.49$ (m/s). This condition is such that it leads to stop-and-go waves in the former algorithm but not in the latter, though all starting with the same initial traffic conditions (all vehicles spaced equally and all with speed of $v^*-1$). A small amount of noise is added in the state evolution for both cases, so as to make the state variables more legible for individual vehicles. 

Figure \ref{SaferTraffic} shows that, right before the onset of a stop-and-go wave (between 345 (s) and 346 (s)), the reaction of the ego to its leading vehicle takes about one second in AS2D$\_$c (Left), leading to a squeezed headway of about 5 (m) (or about 1 (m) bumper-to-bumper distance). In comparison, the actions of the ego and its leading vehicles are nearly synchronized in IAS2D$\_$c with $T=2$ (Right), such that the traffic moves towards a free-flow fixed point with a much safer headway of more than 7 (m) (or 3 (m) bumper-to-bumper) along the way.

\subsubsection{No need for long-range communication}
Clearly, as far as coordination is concerned, there is no need to communicate between vehicles that are beyond the range of interaction with one another. This sets a physical upper bound on how far communication is required. Furthermore, in Figure \ref{SmoothDynamics} we have seen that reaching smoother dynamics faster also entails a fast convergence of the $\tau$-loop (or lower value of $T$). Recall from Figure \ref{DescriptionTable} that the farthest communication reach required is $T+1$. Therefore, another practical benefit is that there is no need to communicate with vehicles that are far away. This will in turn make it easier to practically facilitate the explicit coordination mechanisms proposed in this work.

\subsubsection{No need for complex platooning}
In most CACC studies the focus is on coordinating a few vehicles within a platoon. On the other hand, these works do not address how platoons systematically interact with one another, such as how two nearby platoons merge into a bigger platoon, or vice versa. It is also not easy to model traffic-level dynamics with interacting platoons of various sizes. In contrast, there is no need to define platoon in our approach. We always deal with individual vehicles interacting with some neighboring vehicles within each given coordination mechanism. By aggregating all vehicles together, we can handle all traffic dynamics naturally and seamlessly. Equivalently, all vehicles in the traffic system in our modeling framework is treated as a single extended platoon. 

\section{Summary and future work}\label{Summary}
In order to simultaneously improve traffic efficiency and smoothness, we considered a series of coordinated car-following algorithms. Our approach naturally arises from the modeling framework of the Markov games proposed in \cite{FrameworkPaper}, where individuals' driving policy and traffic control can be solved together as a mechanism design. Computationally, we demonstrated that the pertinent optimal solutions at the traffic level, either based on a Nash equilibrium or centralized optimization, can be well approximated via distributed MPC in real time, implementing the iterated best response dynamics. Our numerical experiments showed very promising results that the coordinated traffic flow can reach as high as twice that of human-driven traffic at high vehicle densities while keeping stop-and-go waves nicely suppressed, provided the ideal speed in the utility function is optimized via a mechanism design procedure offline. An important insight is that the coordinated solutions can attenuate non-smooth transients much faster than uncoordinated solutions. Furthermore, the proposed solutions can rely on similar hardware as those used in the standard CACCs and require only short range communications. Lastly, we performed a linear stability analysis and illustrated why linear stability alone is not sufficient in Appendix \ref{BifurcationAppendix}.

Future work may include the following. More realistic simulations can be investigated, with complications of driver heterogeneity, varying vehicle types and capabilities, communication latency and interruption, measurement error for state estimation, execution noise and so on. Before using real vehicles, the feasibility of proposed solutions could be carefully verified using a fleet of robotic RC cars, which can embody many of the realistic elements just mentioned, relatively quickly and cheaply. Since the iterated response dynamics works beyond the simple car-following settings, it is possible to generalize the coordinated solution concepts to other traffic scenarios, such as multi-lane highways, urban intersections and roundabouts. Of course, the research focus will shift from taming stop-and-go waves to reducing general traffic congestion, either induced by bottlenecks or otherwise, when the driving environments become crowded. The mechanism design associated with new explorations is likely to involve multi-dimensional searches in some subspace among important utility parameters. New measures for traffic efficiency and smoothness are likely needed for characterizing the pertinent collective phenomena. Given the encouraging results we have obtained so far, these future exercises may be fruitful.

\appendix
\makeatletter
\renewcommand{\@seccntformat}[1]{\appendixname~\csname the#1\endcsname:\quad}
\makeatother

\section{Utility components}\label{UtilityComponents}
For the car-following setting there are only three components in the utility sum Eq.(\ref{Cumulative}) or Eq.(\ref{g_transformed}). The evolution of the anticipated states are according to Eq.(\ref{AnticipationProcess}), executed in parallel (or $\forall i\in\{1,\cdots,N\}$) with the current condition: $\hat{s}_{i,h=0}=s_{i,t}$. In the following, we utilize the available information to the fullest, in the sense that in knowing $\hat{s}_{i,h}=(\hat{x}_{i,h},\hat{v}_{i,h},\hat{a}_{i,h})$ we also know $\hat{x}_{i,h+1}=\hat{x}_{i,h}+\hat{v}_{i,h}\Delta t$ and $\hat{v}_{i,h+1}=\hat{v}_{i,h}+\hat{a}_{i,h}\Delta t$.

The first per-period utility component is the reward for moving forward
\begin{equation}
U_{i, t}^{(1)}(u_{i,h}|\hat{s}_{i,h},u_{-i,h})=\exp\Big(-\Big(\frac{\hat{v}_{i, h+1}+u_{i,h}\Delta t - v_i^*}{\kappa_i^{(1)} v_i^*}\Big)^{2}\Big)\, ,
\end{equation}
with the $g$-transform given by
\begin{equation*}
g_1\big[\cup_{h=0}^H\,U_{i, t}^{(1)}(u_{i,h}\big|\hat{s}_{i,h},u_{-i,h})\big]
=U_{i, t}^{(1)}(u_{i,h}|\hat{s}_{i,h},u_{-i,h})\big|_{h=0} \text{  and  } w_{i,1}=1 \, .
\end{equation*}
The second is the penalty for moving backward
\begin{equation}
 U_{i, t}^{(2)}(u_{i,h}|\hat{s}_{i,h},u_{-i,h})=\exp\Big(
-\kappa_v^{(2)}\big(\hat{v}_{i, h+1}+u_{i,h}\Delta t 
+\kappa_0^{(2)}\big)\Big)\, ,
\end{equation}
with the $g$-transform given by
\begin{equation*}
g_2\big[\cup_{h=0}^H\,U_{i, t}^{(2)}(u_{i,h}\big|\hat{s}_{i,h},u_{-i,h})\big]
=U_{i, t}^{(2)}(u_{i,h}|\hat{s}_{i,h},u_{-i,h})\big|_{h=0}\text{  and  } w_{i,2}=-1 \, .
\end{equation*}
The third is a penalty for perceived subjective risk of pairwise collision between the ego $i$ and the leading vehicle $j\equiv i+1$, 
(letting $\mathcal{F}(x)\equiv\exp(-x^2-x)$ and $L_i$ the length of vehicle $i$)
\begin{equation}
U_{i, t}^{(3)}(u_{i,h}\big|\hat{s}_{i,h},u_{-i,h})=\left\{
 \begin{array}{ll}
  1, & \Delta x_{i, j, h} \leq 0 \\ [6pt]
  \mathcal{F} \left(\displaystyle\frac{\Delta x_{i,j,h}}{\delta_{i,j,h}}\right), & \Delta x_{i, j, h}>0
\end{array}\right. \, ,
\end{equation}
\begin{equation*}\text{with}\;\;
\begin{cases}
\Delta x_{i,j,h} = (\hat{x}_{j,h+1}+\hat{v}_{j,h+1}\Delta t-L_{j}/2)
-(\hat{x}_{i,h+1}+\hat{v}_{i,h+1}\Delta t+L_{i}/2)\\
\delta_{i,j,h} = \kappa_c^{(3)}+\kappa_v^{(3)}|
\hat{v}_{i,h+1}+u_{i,h}\Delta t|+\kappa_d^{(3)}
\max\{\hat{v}_{i,h+1}+u_{i,h}\Delta t
- \hat{v}_{j,h+1}-u_{j,h}\Delta t,0\}
\end{cases} ,
\end{equation*}
and the $g$-transform is given by
\begin{equation*}
g_3\big[\cup_{h=0}^H\,U_{i, t}^{(3)}(u_{i,h}\big|\hat{s}_{i,h},u_{-i,h})\big]
=\max_{h\in\{0,\cdots,H\}}U_{i, t}^{(3)}(u_{i,h}\big|\hat{s}_{i,h},u_{-i,h})\, .
\end{equation*}
The weights are chosen as $w_3^g=-10$ for $g$-transformed and $w_3^c=-20$ for cumulative. 

Parameter values, taken from \cite{BifurcationPaper}, are as follows: $C=314$ (m), $N$ varies from 24 to 42, $\Delta t=1/6$ (s), $L=3.9$ (m), $\gamma=\sqrt{0.7}$, $H=7$, $v^*=10.49$ (m/s) unless explicitly specified, $\kappa^{(1)}=0.7$, $\kappa_v^{(2)}=10$, $\kappa_0^{(2)}=0.25$ (m/s), $\kappa_c^{(3)}=0.6$ (m), $\kappa_v^{(3)}=0.3$ (s), $\kappa_d^{(3)}=1.0$ (s). Control input is limited as $u\in \mathcal{U}\equiv[u_\text{min},u_\text{max}]=[-6,4]$ (m/s$^2$). For concrete implementation of Eq.(\ref{softmax}), the chosen grid has 41 points for $u_i^{(0)}\in[-6,4]$ and 11 points for $u_i^{(1)}\in[-1,1]$, with $\lambda=200$. Even for the most sophisticated cases of IAS2D or CAS2D with $T=2$, the coordinated solutions can be achieved in about 10 ms per time step (ie. roughly one sixteenth of $\Delta t$) for a fleet of $N=30$ vehicles on a regular laptop, so long as the code is properly vectorized.

\section{Linear stability analysis}\label{BifurcationAppendix}
Following the procedure developed in \cite{BifurcationPaper} we can derive the following equation whose $z$-roots determine whether the linear stability of the free-flow fixed point of the discrete-time map defined by the dynamics specified in Eq.(\ref{DMPC}):
$k\in\{0,1,\cdots, N-1\}$
\begin{equation}
(\gamma-z)z\Big[(1-z)\Big(1-z+\Delta t(B_k^v-\Delta t B_k^x)\Big)-(\Delta t)^2\,B_k^x\Big]=0\, ,
\end{equation}
where 
$\left(B_k^x, B_k^v , B_k^a \right)\equiv\sum_{l=-\underline{m}}^{\overline{m}}\left(
 \alpha_k^l\beta^x_{l}, \alpha_k^l\beta^v_{l},\alpha_k^l\beta^a_{l}\right)$ and 
\begin{equation}
\alpha_k^l\equiv\exp\big(i\frac{2\pi k l}{N}\big);\quad
\beta_{l,t}^x\equiv\frac{\partial \bar{u}_{i,t}(s^*_{i,t})}{\partial x_{i+l,t}};\quad
\beta_{l,t}^v\equiv\frac{\partial \bar{u}_{i,t}(s^*_{i,t})}{\partial v_{i+l,t}};\quad
\beta_{l,t}^a\equiv\frac{\partial \bar{u}_{i,t}(s^*_{i,t})}{\partial a_{i+l,t}}\, ,
\end{equation}
where $s^*_{i,t}$ denotes the perceivable state evaluated at the fixed point. The free-flow fixed point of the system is linearly stable when all $z$-roots lie inside the unit circle.

\begin{figure}[!h]
\centering
\includegraphics[width=0.85\textwidth]{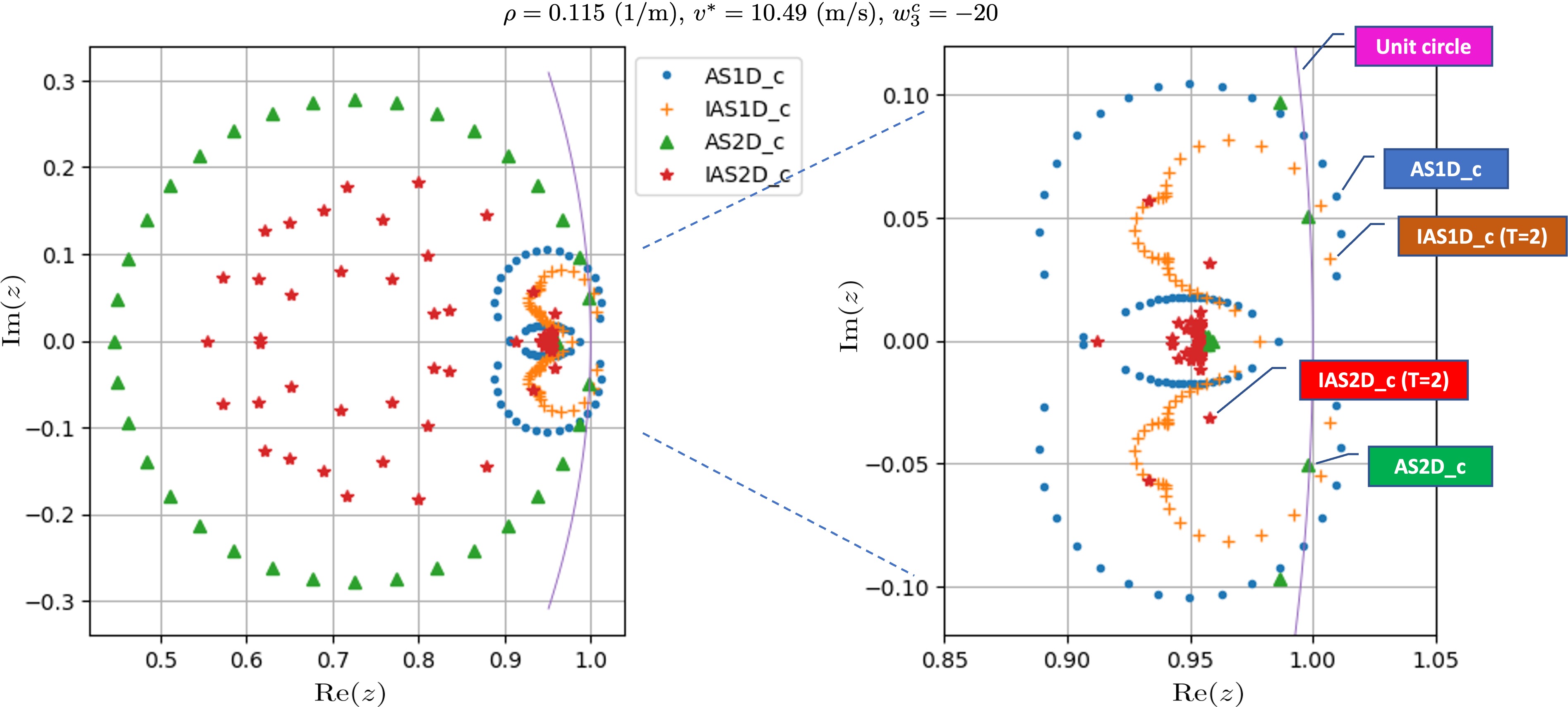}
\caption{$z$-roots for four algorithms: AS1D$\_$c, AS2D$\_$c, IAS1D$\_$c ($T=2$), IAS2D$\_$c ($T=2$), all are with $\rho=0.115$ (1/m) and $v^*=10.49$ (m/s).}
\label{z-roots}
\end{figure}

In Figure \ref{z-roots} we illustrate how $z$-roots distribute for four cases: AS1D$\_$c, AS2D$\_$c, IAS1D$\_$c ($T=2$), IAS2D$\_$c ($T=2$). All are run at $\rho=0.115$ (1/m) and $v^*=10.49$ (m/s). The thin purple line is the unit circle. The $z$-roots for the two algorithms with 1D-grid search (blue and brown) are clearly outside of the unit circle, hence their free-flow fixed points are linearly unstable. The $z$-roots for IAS2D$\_$c (red) all lie inside the unit circle, indicating that its free-flow fixed point is linearly stable. The case of AS2D$\_$c (green) is marginal, as there is a pair of $z$-roots that appear to lie on the unit circle, implying that higher order analysis is required to determine the linear stability of its free-flow fixed point. Of course, we can also use long-term simulations to verify these results.

\begin{figure}[!h]
\centering
\includegraphics[width=0.95\textwidth]{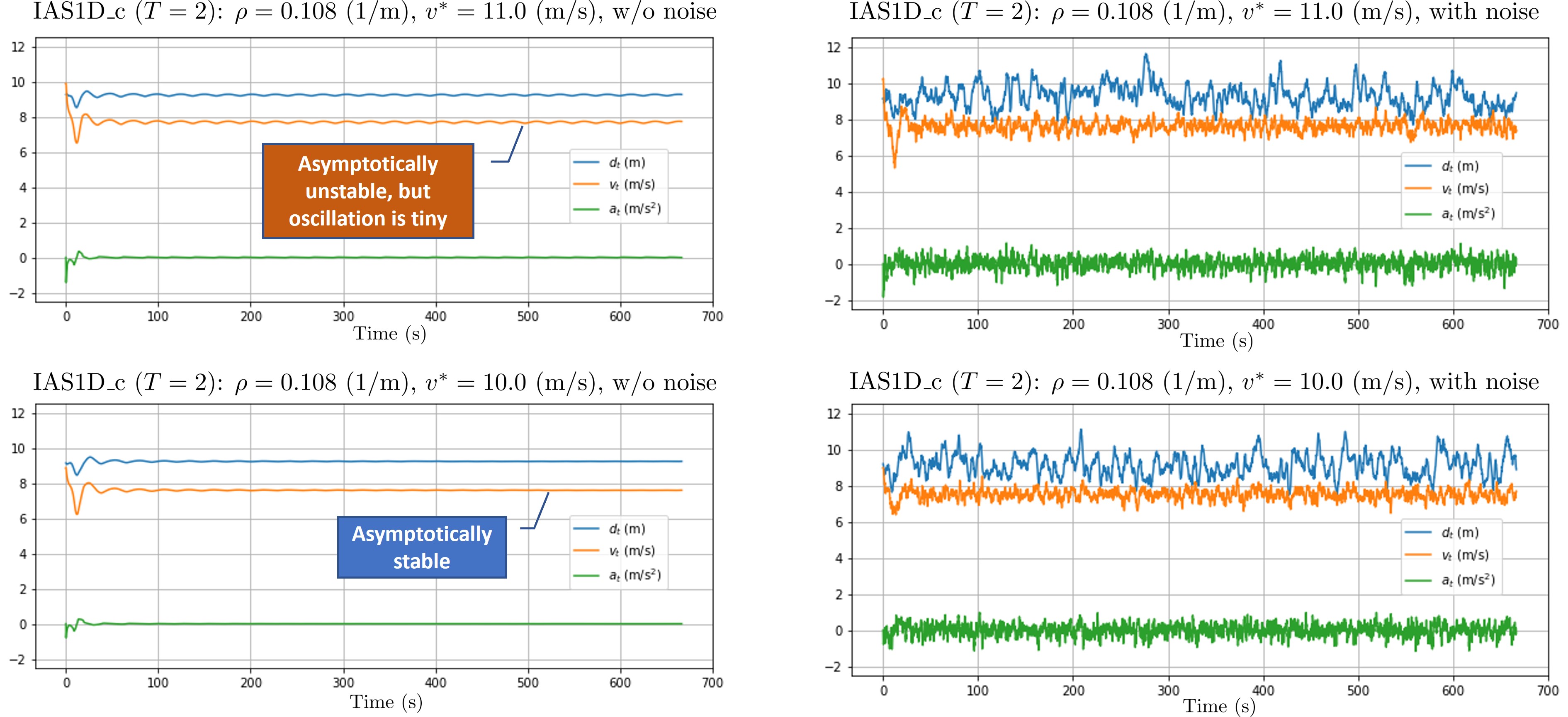}
\caption{An example of a shallow-wave asymptotic solution that is practically indistinguishable from a free-flow fixed-point solution when noises are added.}
\label{ShallowWaves}
\end{figure}

Intriguingly, the linear stability does not tell the full story about traffic stability: 1) In the linearly stable regime: stop-and-go waves can still show up \cite{Orosz, BifurcationPaper}; 2) In the linearly unstable regime: there exist limit solutions with very shallow waves which are indistinguishable from the free-flow fix point when noise is added, as illustrated by Figure \ref{ShallowWaves}. Furthermore, the right panels of the same figure also demonstrate that these coordinated solutions are stable against adding small noises. 

\bibliographystyle{IEEEtran}
\bibliography{refs}

\end{document}